\documentstyle[twocolumn,aps,psfig]{revtex}
\begin{document} 
\title{Ferromagnetism in a dilute magnetic semiconductor --- \\
Generalized RKKY interaction and spin-wave excitations} 
\author{Avinash Singh,$^1$ Animesh Datta,$^2$ Subrat K. Das,$^1$ and Vijay A. Singh$^1$}
\address{$^1$Department of Physics, Indian Institute of Technology Kanpur - 208016} 
\address{$^2$Department of Electrical Engineering, Indian Institute of Technology Kanpur - 208016}
\maketitle
\begin{abstract} 
Carrier-mediated ferromagnetism in a dilute magnetic semiconductor has been studied 
using i) a single-impurity based generalized RKKY approach which goes beyond linear 
response theory, and ii) a mean-field-plus-spin-fluctuation (MF+SF) approach 
within a (purely fermionic) Hubbard-model representation of the magnetic impurities,
which incorporates dynamical effects associated with finite frequency spin correlations 
in the ordered state. Due to a competition between the magnitude of the 
carrier spin polarization and its oscillation length scale, 
the ferromagnetic spin coupling is found
to be optimized with respect to both hole doping concentration and impurity-carrier 
spin coupling energy $J$ (or equivalently $U$). The ferromagnetic transition temperature 
$T_c$, deteremined within the spin-fluctuation theory, corresponds closely with the
observed $T_c$ values. Positional disorder of magnetic impurities causes 
significant stiffening of the high-energy spin-wave modes.
We also explicitly study the stability/instability of the 
mean-field ferromagnetic state, which highlights the role of competing AF interactions 
causing spin twisting and noncollinear ferromagnetic ordering. 
\end{abstract}
\section{Introduction}
The discovery of ferromagnetism in Mn-doped III-V semiconductors
such as p-type In$_{1-x}$Mn$_x$As,\cite{InMnAs}
and Ga$_{1-x}$Mn$_x$As,\cite{GaMnAs} 
with a highest transition temperature ($T_c$) of 110K
for Mn concentration $x=0.053$,\cite{highTc} 
has led to considerable interest 
in these dilute magnetic semiconductors (DMS). 
The successful search for ferromagnetic ordering above room temperature
in Ga$_{1-x}$Mn$_x$N,\cite{GaMnN1,GaMnN2} 
with a highest reported $T_c$ value of 940K,\cite{GaMnN3}
has added a new dimension to the interest.

Besides their potential applications in semiconductor devices 
such as optical isolators, magnetic sensors, 
non-volatile memories seamlessly integrated into 
semiconductor circuits etc., 
and possibilities in photonics and high power electronics,
attention has also been focussed on 
the fundamental mechanism and nature of the ferromagnetic state,
and the possibility of studying new magnetic cooperative phenomena 
such as spin-dependent tunneling, magnetoresistance, spin-dependent light
emission etc. in semiconductor heterostructures arising from
the new (spin) degrees of freedom.

The double-exchange model, 
involving the interaction $- J \vec{S}_i . \vec{\sigma}_i $ between the 
magnetic impurity spin $\vec{S}_i$ and the electron spin $\vec{\sigma}_i$,
has been the starting point in nearly all theoretical studies,
and we first review the emerging physical picture and the 
different approaches employed.

Long range ferromagnetic interaction 
between the $S=5/2$ Mn$^{++}$ ions is mediated,
in the mean-field (Zener model) picture,\cite{mf1,mf2,mf3,mf4,mf5,mf6,mf7}
by a uniform itinerant-carrier spin polarization,
which is caused, in turn, by an effective uniform magnetic field, 
resulting from site-averaging (virtual crystal approximation)
of the local impurity fields.
In the weak-field limit ($xJS << \epsilon_{\rm F}$),
the carrier spin polarization is proportional to 
the Pauli susceptibility $\chi_{\rm P}$,
and the transition temperature ($T_c \sim xJ^2 \chi_{\rm P}$) 
is therefore proportional to the Mn concentration $x$, 
$J^2$, the carrier effective mass $m^*$,
and $N(\epsilon_{\rm F}) \sim p^{1/3}$, where $p$ is the hole concentration.
In DMS, $p$ is a only a small fraction ($f$) of $x$
due to large compensation by As antisite defects.
Therefore, the Fermi energy $\epsilon_{\rm F} \sim W p^{2/3}$ 
itself is quite small compared to the bandwidth $W$,
and hence the weak-field limit is valid only for $x << (W/JS)^3 f^2$. 
Valence band spin splitting comparable in size to the Fermi energy has been
confirmed experimentally.\cite{splitting}

Dynamical correlations in the ordered state have been studied within
a path-integral formulation in which the itinerant carriers are 
integrated out and the effective action for the impurity spins is
expanded up to quadratic order 
(non-interacting spin-wave approximation).\cite{sw1}
In contrast to the MF results, 
the spin stiffness (and hence $T_c$) is independent of $J$ 
and inversely proportional to $m$.
Other approaches incorporating dynamics include 
the dynamical mean field theory,\cite{dmft1,dmft2} 
in which the local charge and spin fluctuations are included
but long-range spin-wave excitations are neglected,
and a RPA-level spin-fluctuation approach in which Mn disorder is treated 
within the coherent potential approximation (CPA).\cite{sw2}

While the positional disorder of Mn ions is not taken into account 
in the virtual crystal approximation (VCA), 
several recent works highlight the importance of 
disorder, both positional and electronic.
Stability of the collinear ferromagnetic state has been investigated 
with randomly distributed Mn ions, and noncollinear ordering is suggested
to be common to these semiconductor systems.\cite{dis1}
Competing (AF) interactions leading to frustration has already been 
evidenced by spin-glass behavior in II-VI DMS.\cite{sg1}
The presence of large compensation due to As antisite defects 
implies substantial electronic disorder as well, 
and the sensitivity of $T_{\rm C}$, magnetization $M$, transport 
and spin-wave spectrum to disorder 
has been investigated.\cite{dis2,dis3,dis4,dis5,dis6}
Monte Carlo simulations have also been used to study  
disorder effects on magnetic ordering,\cite{mc1,mc2,mc3}
and dynamical and transport properties;\cite{mc4}
the background fermions determine the spin interactions 
and hence the nature of the spin ordering, 
which in turn affects the fermionic states. 
Ab-Initio methods\cite{ab1,ab2,ab3,ab4,ab5} have also been recently employed.

An alternative mechanism for the ferromagnetic coupling
between impurity spins involves the hole-mediated RKKY interaction.\cite{highTc} 
The RKKY theory has been extended for various dimensionality structures, 
including effect of potential scattering through carrier mean-free path, 
indicating enhancement of ferromagnetic interaction by disorder 
in low dimensions.\cite{mf1}
Exchange and correlation has been shown to slightly enhance $T_{\rm C}$ 
within the RKKY theory.\cite{mf3}
Spin-wave dispersion in the RKKY picture has been compared with the result of 
spin-wave theory in which a uniform impurity-induced polarization 
has been assumed (VCA), resulting in a Zeeman splitting $\Delta$ 
in the carrier bands.\cite{sw1} It was shown that the RKKY-level dispersion 
is incorrect except when $\Delta << E_{\rm F}$. 

The traditional RKKY approach is based on linear response in the weak-field limit 
($J << \epsilon_{\rm F}$), which is not quite valid for the DMS. 
In this paper, we present a generalized RKKY approach which takes into account
the spatial variation of the impurity-induced carrier spin polarization 
beyond linear response theory [section II].
In the generalized RKKY picture, the local magnetic field $\vec{B}_j = J \vec{S}_j $ 
of a magnetic impurity at site $j$ polarizes the electrons locally, 
and the mobile band electrons spread this magnetic polarization in a 
characteristic manner: $\vec{m}_i = \chi_{ij}(B)\vec{B}_j$,
where $\chi_{ij}(B)$ represents the generalized magnetic response.
The spin $\vec{S}_i$ of another magnetic impurity placed at site $i$ couples to this local electronic magnetization, resulting in an effective 
generalized RKKY spin coupling $J^2 \chi_{ij}(J)\vec{S}_i.\vec{S}_j$. 

We find several interesting competing processes which 
limit the growth of spin couplings. 
As the RKKY response involves a particle-hole process, 
it vanishes for a filled (valence) band 
and grows with increasing hole concentration $p$. 
While the spin coupling is therefore expected to strengthen with $p$, 
a competing process involving the length scale sets in,
which limits the growth of the spin coupling and therefore of 
the ferromagnetic transition temperature $T_c$.
The Fermi wavelength $\lambda_{\rm F}= 2\pi/k_{\rm F}$,
which sets the RKKY oscillation length scale, decreases with hole doping,
and therefore the spin coupling between two magnetic impurities 
at a fixed separation goes through a maximum as a function of  
hole concentration [section III]. 

We find a similar optimization in the spin coupling 
as a function of the impurity field strength $B$.
By going beyond linear response theory, and examining the 
generalized RKKY response for a fixed hole concentration,
we find that the RKKY oscillation becomes more rapid with increasing polarizing field.
Therefore, for a fixed separation between two impurity spins, 
the spin coupling initially increases like $J^2$ as expected,
but then crosses over and eventually changes sign, resulting in frustration [section III]. 
The non-linear magnetic response thus brings out another 
limitation in the ferromagnetic spin coupling.

In order to determine the extent to which the magnetization response 
of a single impurity determines the macroscopic magnetic properties
of the DMS, we have also considered a finite concentration of magnetic
impurities distributed on a finite-size lattice [section IV].  
Using a novel Hubbard-$U$ representation for the magnetic impurities in a DMS, 
we have studied the collective magnetic response in the ferromagnetic state
within a mean-field-plus-spin-fluctuation (MF+SF) approach. 
Treating the disorder aspects of the Mn-impurity system exactly,
and electron correlation effects within the random phase approximation (RPA), 
our numerical analysis yields the spin stiffness from the 
low-lying collective (spin-wave) excitations [section VI],
which have a fundamental bearing on the ferromagnetic transition temperature $T_c$.
Our approach also allows for a quantitative study of the 
stability/instability of the Hartree-Fock (mean-field) ferromagnetic state [section V], 
highlighting the presence of competing interactions.
The Anderson Hamiltonian, with a hybridization term $V_{\rm pd}$ 
between band fermions and the magnetic impurity orbital, 
has also been recently studied to obtain the ferromagnetic coupling 
between two magnetic impurities.\cite{ivanov}

\section{Magnetic impurity in a host}
We consider a single-band spin-fermion lattice model 
\begin{equation}
H = \sum_{{\bf k},\sigma} 
\epsilon_{\bf k}
a_{{\bf k},\sigma}^{\dagger} a_{{\bf k},\sigma}
- \sum_{i} J_i \vec{S}_i . \vec{\sigma}_i \; ,
\end{equation}
with a double-exchange interaction between the 
magnetic impurity spin $\vec{S}_i$ and the electron spin $\vec{\sigma}_i$
at the impurity site $i$. 
The host (valence band) dispersion $\epsilon_{\bf k}$ is taken to be 
parabolic for small $k$ (top of the band at $k=0$), 
the $k^2$ coefficient determining the inverse carrier mass $m^*$.
As the added holes go in long-wavelength states,
the small-$k$ particle-hole processes near the Fermi energy are dominant, 
and therefore other details of the energy band are expected to be 
relatively unimportant. 

\subsection{Host Green's function}
We consider an isotropic energy-band dispersion 
\begin{equation}
\epsilon_k = \frac{W}{2} \cos ka \; 
\end{equation}
in three dimensions, 
with the wavevector magnitude extending upto $\pi/a$.
This dispersion incorporates the desired features, 
and yields a finite bandwidth without introducing sharp cutoffs. 
We choose length and energy units such that the lattice spacing $a=1$ 
and the bandwidth $W=1$.  
The advanced Green's function for the host is obtained as
\begin{figure}
\vspace*{-70mm}
\hspace*{-38mm}
\psfig{file=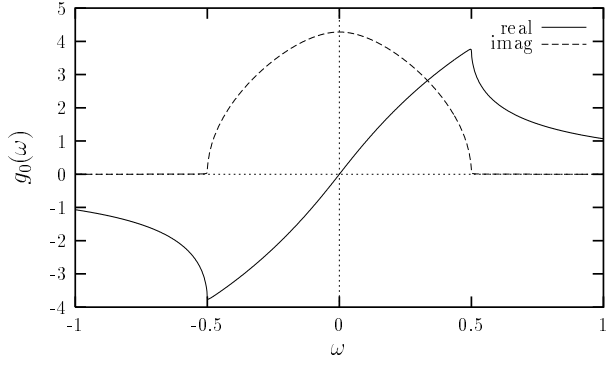,width=135mm,angle=0}
\vspace{-75mm}
\caption{Real and imaginary parts of the local host Green's function.}
\end{figure}

\begin{eqnarray}
g_{ij}(\omega) &=& \frac{1}{N} \sum_{\bf k}  
\frac{e^{i{\bf k}.({\bf r}_i - {\bf r}_j)}}
{\omega-\epsilon_{\bf k} -i\eta} \nonumber \\
g_r (\omega)&=& 
\int_0 ^\pi \beta(k) dk \int_0 ^\pi \frac{1}{2} \sin \theta \; d\theta 
\frac{e^{ikr\cos \theta}}
{\omega-\epsilon_{k} -i\eta} \nonumber \\
&=& 
\int_0 ^\pi \beta(k) dk 
\frac{1}{\omega-\epsilon_{k} -i\eta} 
\frac{\sin kr}{kr}
\end{eqnarray}
Here $\beta(k)$ is a $k$-space density of states, 
and for simplicity we choose a symmetric form
\begin{eqnarray}
\beta(k) &=& ak^2 - bk^4 \hspace{1in}  (0 \le k \le \pi/2) \nonumber \\
         &=& a(k-\pi)^2 - b(k-\pi)^4 \hspace{.29in}  (\pi/2 \le k \le \pi) \; ,
\end{eqnarray}
so that the usual three-dimensional $k^2$ form is recovered for 
states near both the lower and upper band edges at 
$k=\pi$ and $k=0$, respectively. 
We choose $b=2a/\pi^2$, so that $\beta(k)$ is smooth at $k=\pi/2$
(the slope $d\beta/dk = 0$),
and an overal normalization $a=120/7\pi^3$
so that the sum over states in the band $\int_0 ^\pi \beta(k) dk =1$.

The above choice yields a symmetric band 
with a nearly semi-elliptical density of states, as seen in Fig. 1, 
showing the real and imaginary parts of the local host Green's function $g_0(\omega)$.
Near the band edges, the real-part magnitude has a finite maximum
and the imaginary-part has a square-root behaviour, as expected for the 
three-dimensional system. 
The band filling is shown in Fig. 2 as a function of the Fermi energy. 
\subsection{Magnetic response}
We consider the impurity spin in the classical limit
($J_i \vec{S}_i \rightarrow J_i \langle \vec{S}_i \rangle = B_i \hat{z}$),
and examine the magnetic response of electrons in a 
nearly filled band due to the magnetic coupling 
$- \sum_{i}  \vec{\sigma}_i . \vec{B}_i $,
for an arbitrary strength of 
the impurity-induced local magnetic field $\vec{B}_i$.

For a single magnetic impurity at site $j$, 
the electronic Green's function $G$ is exactly obtained in terms of the 
host Green's function $g$ as 

\begin{figure}
\vspace*{-70mm}
\hspace*{-38mm}
\psfig{file=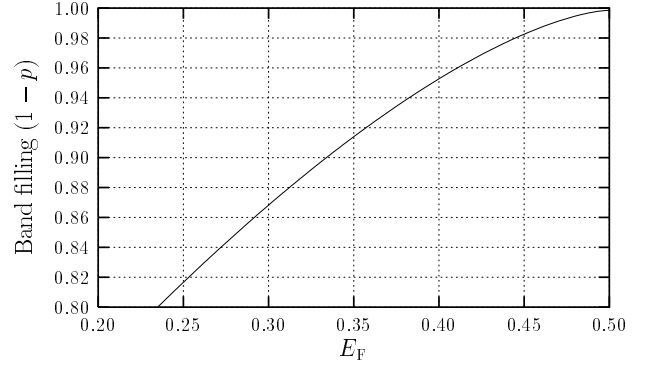,width=135mm,angle=0}
\vspace{-75mm}
\caption{Fermi-energy dependence of the band filling.}
\end{figure}

\begin{equation}
G_{ii}^\sigma (\omega) = 
g_{ii}(\omega) + 
g_{ij}(\omega)\left [ \frac{-\sigma B_j }
{1+\sigma B_j g_0 (\omega)} \right ] 
g_{ji}(\omega) \; ,
\end{equation}
where $g_0 \equiv g_{jj}$ is the local host Green's function.
The resulting local magnetization $m_i$ at site $i$ is then obtained as 
\begin{equation}
m_i = 
\int_{-\infty} ^{\omega_{\rm F}} \frac{d\omega}{\pi} 
{\rm Im}
[G_{ii}^\uparrow (\omega) - G_{ii}^\downarrow (\omega)] \; ,
\end{equation}
where 
\begin{equation}
G_{ii}^\uparrow (\omega) - G_{ii}^\downarrow (\omega) = 
g_{ij}(\omega) \; \Delta T_j \; g_{ji}(\omega) \; ,
\end{equation}
in terms of the $T$-matrix difference 
\begin{equation}
\Delta T_j \equiv 
T_j ^\uparrow - T_j ^\downarrow =
\left [ \frac{-2 B_j }{1 - B_j ^2 g_0 ^2(\omega)} \right ] 
\; .
\end{equation}

Defining a field-dependent generalized magnetic response function 
$\chi_{ij}(B)$ through the relation
\begin{equation}
m_i = \chi_{ij}(B) B_j \;,
\end{equation}
Eqs. (6), (7), (8) yield
\begin{equation}
\chi_{ij}(B)= 
\int_{-\infty} ^{\omega_{\rm F}} \frac{d\omega}{\pi} 
{\rm Im}
\left [
g_{ij}(\omega) 
\left ( \frac{-2}{1 - B^2 g_0 ^2(\omega)} \right )
g_{ji}(\omega) \right ] \; .
\end{equation}

\section{Effective spin couplings}
Another impurity spin $\vec{S}_i$ 
placed at site $i$ will couple with the local magnetization $m_i$
produced by the local field of the spin $\vec{S}_j$ at site $j$,
resulting in an effective interaction between the two spins given by
\begin{equation}
H_{\rm spin}(J)= -J_i J_j \chi_{ij}(B=JS) \vec{S}_i . \vec{S}_j  \; .
\end{equation}

\begin{figure}
\vspace*{-70mm}
\hspace*{-38mm}
\psfig{file=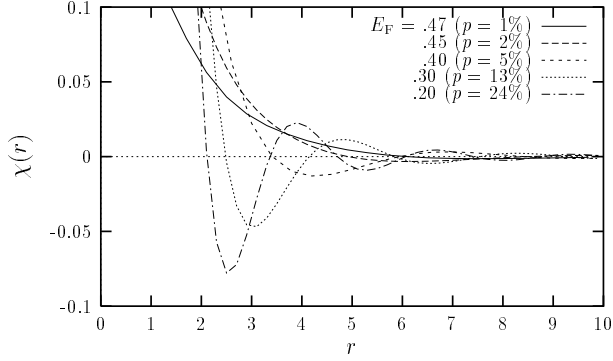,width=135mm,angle=0}
\vspace{-75mm}
\caption{Magnetic susceptibility $\chi(r)$ for different hole doping concentrations.}
\end{figure}

\begin{figure}
\vspace*{-70mm}
\hspace*{-28mm}
\psfig{file=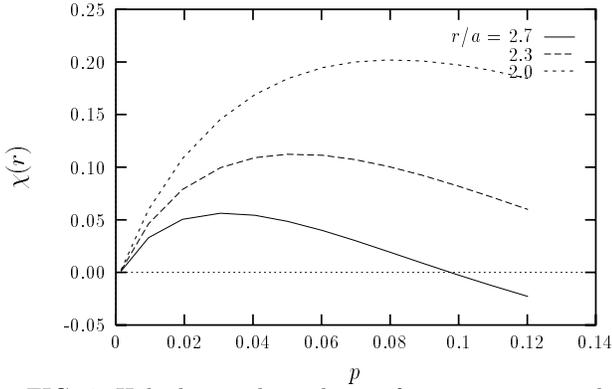,width=135mm,angle=0}
\vspace{-75mm}
\caption{Hole doping dependence of magnetic susceptibility $\chi(r)$ for different $r$ 
(corresponding to average Mn-Mn separations in a cubic host lattice with 
$5\%$, $8\%$, and $12.5\%$ Mn impurity concentration).}
\end{figure}

\subsection{Weak-coupling limit: RKKY interaction}
When the $B^2$ term in Eq. (10) can be neglected
(valid for $B << W$), one obtains a linear response
\begin{equation}
m_i = \chi_{ij} B_j \;,
\end{equation}
where the magnetic susceptibility $\chi_{ij}$ 
\begin{eqnarray}
\chi_{ij} & = & 
-2 \int_{-\infty} ^{\infty} \frac{d\omega}{\pi}
\;{\rm Im} [g_{ij}(\omega) g_{ji}(\omega)] \nonumber \\
& = & 
4\sum_{\epsilon_l < \epsilon_{\rm F} }
\sum_{\epsilon_m > \epsilon_{\rm F} }
\frac{\phi_l^i \phi_l ^{j*} \phi_m^j \phi_m ^{i*} }
{\epsilon_m - \epsilon_l}  \; ,
\end{eqnarray}
yields the standard oscillating RKKY interaction 
\begin{equation}
H_{\rm RKKY}= -J_i J_j \chi_{ij} \vec{S}_i . \vec{S}_j \; .
\end{equation}

The behaviour of $\chi_{ij}$, as a function of the separation $r$ betwen the
two sites $i$ and $j$, shows that the oscillation becomes
more rapid with doping (Fig. 3), as expected from the decreasing
Fermi wavelength $\lambda_{\rm F}=2\pi/k_{\rm F}$. 
Qualitatively similar results were obtained for a parabolic 
energy dispersion $\epsilon_{\bf k} \sim k^2$ with a finite bandwidth cutoff.

For a fixed separation $r/a=(1/x)^{1/3}$, 
corresponding to the average Mn-Mn distance in a cubic lattice with 
Mn concentration $x$, 
the behaviour of $\chi(r)$ is shown in Fig. 4 as a function of band filling. 
The ferromagnetic coupling peaks at hole concentrations about 
$0.6$ times the Mn impurity concentration.

\subsection{Generalized magnetic response}
It appears that the conventional RKKY picture based on the weak-coupling limit $(B<<W)$ 
cannot provide a good description of the interaction between 
Mn impurities in Ga$_{1-x}$Mn$_x$As. 
Core-level photoemission\cite{jval1} yields $J \sim 1$ eV, 
which is comparable to the host bandwidth of $W \approx 2$ eV
for the heavy hole band.\cite{bandwidth}  
It is therefore essential to go beyond the linear-response regime,
and for $B \sim W$ we find that there are additional contributions 
in the generalized magnetic response function $\chi_{ij}(B)$, which 
qualitatively modify the nature of the magnetic response and spin couplings.
\subsubsection{Impurity-state contribution}
For $\omega$ outside the band ($|\omega| > W/2$), 
the T-matrix difference in Eq. (10) has imaginary terms of the type $\delta(\omega-\omega^*)$,
arising from the two poles 
\begin{equation}
1 \pm B g_0(\omega^*) = 0 \; ,
\end{equation}
corresponding to a spin-$\uparrow$ impurity state
at $\omega^* _\uparrow $ (below the lower band edge),
and a spin-$\downarrow$ impurity state
at $\omega^* _\downarrow $ 
(above the upper band edge). 
In three dimensions, $g_0(\omega)$ has a finite maximum at the band edges,
and therefore impurity states are formed only when 
$B$ exceeds a threshold strength $B^*$. 

By expanding $g_0(\omega)$ near $\omega^* _\uparrow$, 
and expressing $T_j ^\uparrow$ as a simple pole, 
the impurity-induced correction is given by 
\begin{equation}
G_{ii}^\uparrow - g_{ii} = g_{ij} T_j ^\uparrow g_{ji} 
= \frac{|\varphi_i ^\uparrow |^2 }
{\omega - \omega_\uparrow ^* -i\eta} \; ,
\end{equation}
where the impurity-state wavefunction 
$\varphi_i ^\uparrow $ is given by 
\begin{equation}
\varphi_i ^\uparrow = \frac{g_{ij}(\omega=\omega_\uparrow ^*)}
{\sqrt{-dg_0/d\omega |_{\omega=\omega_\uparrow ^*}}} \; .
\end{equation}
For any finite doping, 
only the spin-$\uparrow$ impurity state is occupied,
and the impurity-state contribution to the local magnetization 
is therefore simply obtained as
\begin{equation}
m_i ^* = |\varphi_i |^2 \;.
\end{equation}
With  increasing $B$, the impurity-state wavefunction becomes more localized, 
and $\varphi_i \rightarrow \delta_{ij}$
as $B \rightarrow \infty$.

\begin{figure}
\vspace*{-70mm}
\hspace*{-38mm}
\psfig{file=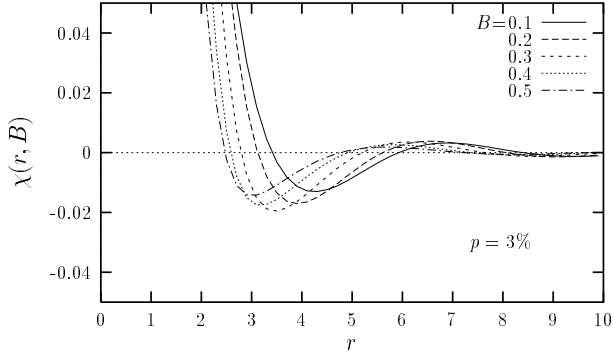,width=135mm,angle=0}
\vspace{-75mm}
\caption{The $r$-dependence of the generalized magnetic response 
$\chi(r,B)$, for different field strengths.}
\end{figure}

\begin{figure}
\vspace*{-70mm}
\hspace*{-38mm}
\psfig{file=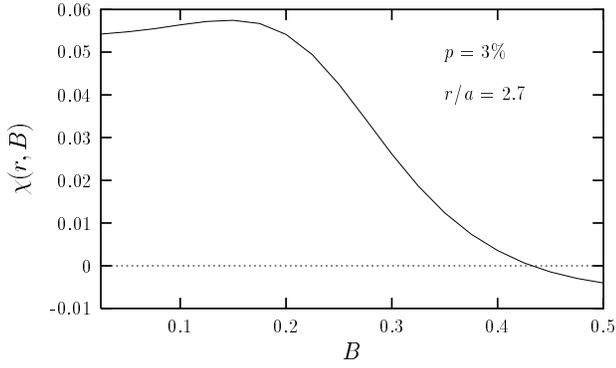,width=135mm,angle=0}
\vspace{-75mm}
\caption{The field-strength dependence of the generalized magnetic response 
$\chi(r,B)$, for a fixed hole concentration and Mn-Mn distance.}
\end{figure}

\subsubsection{Band contribution}
The other contributions to the imaginary part in Eq. (10)
are from within the band ($|\omega| < W/2$), 
and involve the real (imaginary) part of $\Delta T_j(\omega)$ 
and imaginary (real) part of $g_{ij}(\omega) g_{ji}(\omega)$. 

Including both the band and impurity contributions,
the generalized magnetic response $\chi(r,B)$ evaluated from Eq. (10) 
is shown in Fig. 5 for different field strengths;
the lowest-field case ($B=0.1$) provides the RKKY response 
($B\rightarrow 0$), for comparison. 
The length scale at which the first crossover from ferromagnetic to antiferromagnetic 
coupling takes place is seen to decrease with increasing $B$.
Fig. 6 shows the generalized magnetic response $\chi(r,B)$,
for a fixed hole concentration and Mn-Mn distance.
For small $B=JS$, the response is essentially constant (linear response), 
and in this regime the generalized RKKY interaction energy $B^2 \chi(r,B)$ 
grows like $J^2$, as in the mean-field and conventional RKKY pictures. 
However, the sharp suppression in the generalized magnetic response for $B/W > 0.2$
limits this growth and leads to a peak, 
which is seen to shift to higher $B$ values with decreasing Mn-Mn separation (Fig. 7).
This effect
\begin{figure}
\vspace*{-70mm}
\hspace*{-28mm}
\psfig{file=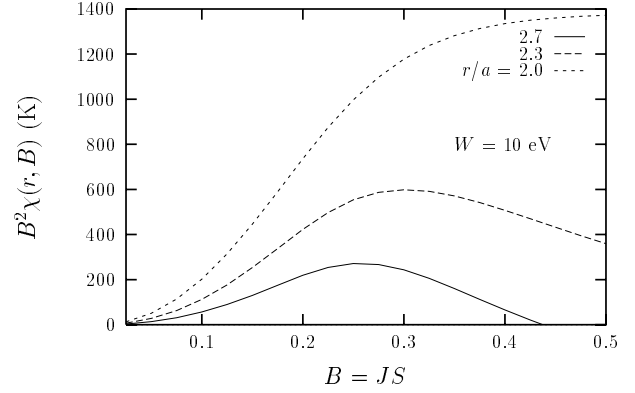,width=135mm,angle=0}
\vspace{-70mm}
\caption{The generalized RKKY interaction energy $B^2 \chi(r,B)$ as a function
of the local field strength $B$, for different Mn-Mn distances. 
The host bandwidth is nominally taken as 10 eV.}
\end{figure}

\noindent
significantly increases the spin coupling 
in a higher Mn concentration system (such as GaMnN)
beyond the factor expected from the spin response (Fig. 4).

As the effective carrier mass $m^*$ scales like the inverse bandwidth $1/W$,
the $m^*$ dependence of the generalized magnetic response $\chi(r,B)$ 
can be directly deduced from Fig. 6,
showing the $B/W \propto m^*$ dependence for a fixed $B=JS$.
The magnetic response in the (fixed) unit of $1/B$ is obtained by 
multiplying $\chi(r,B)$ in Fig. 6 (in unit of $1/W$) by $B/W$. 
This yields a linear $m^*$ dependence of $\chi(r,B)$ 
(and hence the spin coupling energy and $T_c$) for low effective mass
and then a sharp suppression with increasing $m^*$.
Sublinear dependence of $T_c$ at large $m^*$ has also been reported  
in Monte Carlo studies.\cite{mc1} 

For a nominal host bandwidth of 10eV (with 1 eV $\approx 10^4$ K),
the peak interaction energies are about 250 K, 600 K and 1400 K
for Mn concentrations of $5\%$, $8\%$, and $12.5\%$, and   
hole concentrations of $3\%$, $5\%$, and $8\%$, respectively [Fig. 7].
From this spin interaction energy $(JS)^2 \chi$,
the ferromagnetic transition temperature 
$T_c$ can be estimated within the spin-fluctuation theory.
For a nearest-neighbour quantum Heisenberg model (interaction energy 
$\cal J$) on a hypercubic lattice (coordination number $z$),
the transition temperature is given by $T_c = T_c^{\rm MF}/f_{\rm sf}$,
somewhat lower than the mean-field value   
$T_c^{\rm MF}={\cal J}S(S+1) z/3 $.\cite{doniach}
Here $f_{\rm sf} = (1/N)\sum_{\bf k}(1-\gamma_{\bf k})^{-1} \gtrsim 1 $
is a geometrical spin-fluctuation factor, 
where $\gamma_{\bf k} \equiv (\cos k_x + \cos k_y + \cos k_z)/3$
in three dimensions.

As the effective RKKY interaction term between two spins
is $J^2 \chi_{ij} \vec{S}_i. \vec{S}_j$, 
we take ${\cal J}=J^2 \chi$ and obtain
$T_c \approx 2J^2 \chi S(S+1) $ for $z=6$. 
Taking a realistic bandwidth of $W=2$eV 
for the heavy valence band,\cite{bandwidth}
the peak energy in Fig. 7 translates to a peak $T_c$ 
of about 150 K and 850 K for $5\%$ and $12.5\%$ Mn concentrations,  
quite close to the observed highest $T_c$ values for 
Ga$_{1-x}$Mn$_x$As and Ga$_{1-x}$Mn$_x$N.

\section{Hubbard-$U$ representation of magnetic impurities}
We now consider a (purely fermionic) Hubbard-model representation 
for the randomly distributed magnetic impurities on a cubic lattice:
\begin{eqnarray}
H &=& t \sum_{<ij>\sigma} \left (\hat{a}_{i\sigma}^{\dagger}\hat{a}_{j\sigma}
+ {\rm h.c.} \right ) 
+ t' \sum_{<Ij>\sigma} \left 
(\hat{a}_{I\sigma}^{\dagger}\hat{a}_{j\sigma}
+ {\rm h.c.} \right ) \nonumber \\
&+& \epsilon_d \sum_{I,\sigma} \hat{a}_{I\sigma}^\dagger \hat{a}_{I\sigma}
+ U \sum_{I} 
\left (\hat{n}_{I\uparrow}- n_I \right )
\left (\hat{n}_{I\downarrow}- n_I \right ) \;,
\end{eqnarray}
where I refers to the impurity sites, 
$\epsilon_d$ is the impurity on-site energy
and $n_I = \langle \hat{n}_{I\uparrow} + \hat{n}_{I\downarrow} \rangle /2$ 
is the spin-averaged impurity charge density. 
Higher spin magnetic impurities, such as the $S=5/2$ Mn impurities in 
$\rm Ga_{1-x}Mn_x As$, can be realistically represented within a generalized
Hubbard model representation involving multiple orbitals 
and different interaction processes 
(direct and exchange type, with respect to orbital indices).\cite{magimp}
For simplicity, we have taken the same hopping ($t'=t=1$) 
between the host-host and host-impurity nearest-neighbour pairs of sites.
The energy-scale origin is set so that the host on-site energy is zero,
and we take the impurity level to lie at the top of the host band ($\epsilon_d =6$). 
The form of the Hubbard interaction term is such that in the Hartree-Fock 
approximation it reduces to the double-exchange term.

\subsection{Hartree-Fock ferromagnetic state}
In the Hartree-Fock (mean-field) approximation, the interaction term reduces to
a magnetic coupling of the electron
to the local mean (magnetic) field $\vec{\Delta}_I$:
\begin{equation}
H_{\rm int}^{\rm HF} = - \sum_{I} 
\vec{\sigma}_I . \vec{\Delta}_I  \; ,
\end{equation} 
where the electronic spin operator 
$\vec{\sigma}_I =  \Psi_I ^\dagger [\vec{\sigma}] \Psi_I $
in terms of the spinor  
$\Psi_I =  \left ( \begin{array}{l} 
\hat{a}_{I\uparrow} \\ 
\hat{a}_{I\downarrow} \end{array} \right ) $,
and the mean field $\vec{\Delta}_I$
is self-consistently determined from the ground-state expectation value: 
\begin{equation}
2\vec{\Delta}_I = U \langle \vec{\sigma}_I \rangle \; .
\end{equation}
Thus, in the classical (Hartree-Fock) limit, the interaction term  
reduces to the corresponding form of the double-exchange term, 
with the mean field $\vec{\Delta}_I$ representing
the impurity-induced local magnetic field $\vec{B}_I$.

Starting with an initial uniform mean field $\vec{\Delta}_I = \hat{z} \Delta$,
the mean-field (MF) Hamiltonian is numerically diagonalized for a finite lattice
to obtain the fermion eigenfunctions $\phi_{l\sigma}$ and eigenvalues $E_{l\sigma}$.
The spin densities $n_{I\sigma}=\sum_{E_{l\sigma} < E_{\rm F}} 
(\phi_{l\sigma}^I)^2 $
yield the new local mean fields $\Delta_I = U (n_{I\uparrow}-n_{I\downarrow})/2$,
which are then used to update 

\begin{figure}
\vspace*{-70mm}
\hspace*{-28mm}
\psfig{file=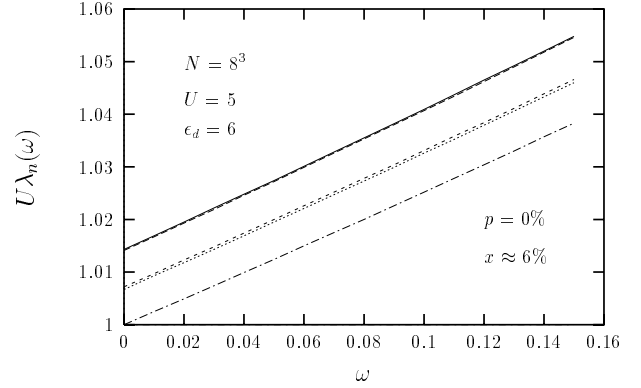,width=135mm,angle=0}
\vspace{-75mm}
\caption{Instability of the undoped (HF) ferromagnetic state.
The Goldstone mode ($U\lambda_G = 1$) corresponds to the 
minimum eigenvalue of the $[\chi^0 (\omega=0)]$ matrix,
indicating maximal instability.}
\end{figure}

\begin{figure}
\vspace*{-70mm}
\hspace*{-28mm}
\psfig{file=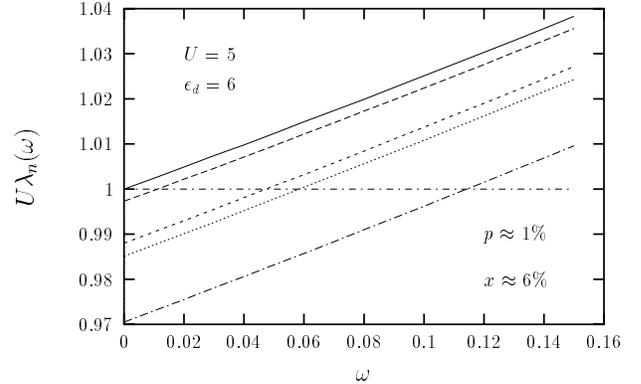,width=135mm,angle=0}
\vspace{-75mm}
\caption{Stabilization of the ferromagnetic state with hole doping;
the Goldstone mode now corresponds to the maximum eigenvalue.}
\end{figure}
\noindent
the MF Hamiltonian, 
and this procedure is iterated till self-consistency is achieved.

\subsection{Stability of the HF state}
The self-consistent, HF ferromagnetic state, 
with all local moments aligned in the same symmetry-breaking 
direction, does not necessarily represent a stable (lowest-energy) state. 
This is because the HF state really represents an energy extremum,
which may correspond to a saddle point having local energy minimum and maximum 
along different directions in the order-parameter space.
The stability of the HF state
with respect to transverse perturbations in the order parameter
is indicated by the maximum eigenvalue
$\lambda_{\rm max}$ of the $[\chi^0 (\omega=0)]$ matrix (Eq. 24).\cite{doped}
The HF state is stable if $U\lambda_{\rm max} = 1$,
correspondng to the massless Goldstone mode, 
representing a rigid rotation of the ordering direction.
Instability is indicated if $U\lambda_{\rm max} > 1$, 
signalling a growth of transverse perturbations about the HF state,
which can also be interpreted as negative-energy bosonic modes.

Figure 8 shows some of the eigenvalues 
of the $[\chi^0 (\omega)]$ matrix (including the minimum and maximum)
for the undoped HF ferromagnetic state of an $8^3$ system, 
with a semi-ordered arrangement of 32 magnetic impurities (see section V for details).
For $\omega=0$, the Goldstone mode ($U\lambda_G = 1$) is seen to 
correspond to the {\em lowest eigenvalue}, 
indicating maximal instability of the ferromagnetic state. 
The structure of the eigenvector corresponding to the maximum eigenvalue 
indicates a tendency towards AF ordering of the impurity spins. 
The AF coupling arises from the exchange interaction
$J' \sim t'^2 /U$ due to the effective hopping $t'$ 
(associated with impurity-band formation) between impurity sites.  
In the absence of the hole-induced (RKKY) ferromagnetic coupling, 
this exchange interaction dominates and favours AF ordering of impurity spins.
With hole doping, the ferromagnetic state gets stabilized, 
and the Goldstone mode now corresponds to the maximum eigenvalue (Fig. 9).

\subsection{Transverse spin fluctuations}
Transverse spin fluctuations are gapless, low-energy excitations in 
the broken-symmetry state of magnetic systems possessing
continuous spin-rotational symmetry. 
Therefore, at low temperatures they play an important
role in diverse macroscopic properties
such as existence of long-range order, 
magnitude and temperature dependence of the order parameter,
magnetic transition temperatures, spin correlations etc.

We study the time-ordered, spin-wave propagator 
involving the spin-lowering ($S_i ^-$) and spin-raising ($S_j ^+$) operators 
at sites $i$ and $j$:
\begin{equation}
\chi_{ij}(t-t') =  i
\langle \Psi_{\rm G} | T [ S_i ^- (t) S_j ^+ (t')]|\Psi_{\rm G}\rangle \; .
\end{equation}
At the RPA level, the spin-wave propagator in frequency space is given by
\begin{equation}
[\chi^{-+}(\omega)]=\frac{[\chi^0(\omega)]}{1-[U][\chi^0(\omega)]} \;,
\end{equation}
where the zeroth-order, antiparallel-spin particle-hole propagator

\begin{eqnarray}
[\chi^0(\omega)]_{ij} &=& i\int \frac{d\omega'}{2\pi}
G_{ij}^{\uparrow}(\omega')G_{ji}^{\downarrow}(\omega'-\omega) \\
&=& 
\sum_{E_l < E_{\rm F}}
^{E_m > E_{\rm F}}
\left (
\frac
{
\phi_{l\uparrow}^i \phi_{m\downarrow}^i
\phi_{m\downarrow}^j \phi_{l\uparrow}^j
}
{
E_{m\downarrow} - E_{l\uparrow} + \omega
}
+
\frac
{
\phi_{l\downarrow}^i \phi_{m\uparrow}^i
\phi_{m\uparrow}^j \phi_{l\downarrow}^j
}
{
E_{m\uparrow}-E_{l\downarrow} - \omega
}
\right ) \nonumber 
\end{eqnarray}
is evaluated using the eigenvalues $E_{l\sigma}$ and eigenvectors $\phi_{l\sigma}$
in the self-consistent, broken-symmetry state. 
In Eq. (23), the diagonal interaction matrix $[U]_{ii}=U \delta_{iI}$
has non-vanishing elements only at the magnetic impurity sites.
For site-dependent interactions, it is convenient to 
recast Eq. (23) using simple matrix manipulations:
\begin{equation}
[\chi^{-+}(\omega)]=
\frac{1}{[A(\omega)]} - \frac{1}{[U]} \; ,
\end{equation}
where $[A(\omega)]=[U] - [U][\chi^0 (\omega)][U]$ is a symmetric matrix,
having non-vanishing matrix elements only in the reduced impurity basis:
\begin{equation}
[A(\omega)]_{IJ}=U(1 - U[\chi^0 (\omega)]_{IJ}) \; .
\end{equation}

Spin-wave modes, 
represented by the poles in the propagator $[\chi^{-+}(\omega)]$, 
are hence given by the poles of the matrix $[A(\omega)]_{IJ}$,
as $[U]$ is non-singular. In 
terms of the eigenvalues $\lambda_n$ and eigenvectors $\varphi _n $ of the 
$[\chi^0 (\omega)]_{IJ}$ matrix,
the spin-wave energies $\omega_n$ are therefore given by 
\begin{equation}
1-U\lambda_n (\omega_n) =0 \; .
\end{equation}

\section{Spin-wave energy}
The spin couplings and stiffness in the ferromagnetic state 
can be determined from the spin-wave energies.
To see how the magnitude and sign of the spin couplings  
depend on impurity separation, we have considered 
several impurity arrangements with different numbers ($N_{\rm imp}$) 
of magnetic impurities in a cubic host lattice with $N=8^3$ sites.
These arrangements include: 
(i) an ordered impurity arrangement of 64 impurities ($x=1/8$)
on a cubic superlattice with impurity separation $2a$,
(ii) a semi-ordered arrangement of 32 impurities ($x\approx 6\%$), 
with same NN impurity separation ($2a$) in the z direction 
but greater in-plane separation ($\sqrt{8}a$),
and (iii) a disordered arrangement of 30 impurities ($x\approx 6\%$)
with NN separations ranging between $2a$ and $3a$. 

For the undoped (insulating) state,  
we take $N_\uparrow = N$ and $N_\downarrow=N-N_{\rm imp}$;
all spin-$\downarrow$ impurity states 
(pushed up by the local mean field) are then unoccupied,
resulting in local-moment formation.
Hole doping is introduced by reducing $N_\uparrow$,
and band fillings are so chosen that the Fermi energy lies
in gaps between (nearly) degenerate groups of eigenvalues. 

The undoped self-consistent ferromagnetic state 
is found to be maximally unstable, as discussed earlier.
Indeed, the self-consistent antiferromagnetic state is actually found to be stable,
confirming the dominance of the AF spin couplings $J' \sim t'^2 /U$.
With hole doping, the ferromagnetic state is stabilized, 
and the spin-wave energies $\omega_n$ are extracted from the pole 
condition $U\lambda_n (\omega_n) = 1$.
Hole doping dependence of the lowest spin-wave energy  
$\omega_{\rm low}$ is shown in Fig. 10
for the ordered (i) and disordered (ii) impurity arrangements, 
with $U=4$ and $5$, respectively.  
Figure 11 shows the $U$-dependence of the lowest spin-wave energy
for the ordered (i) and disordered (iii) arrangements,
with $N_\uparrow = 482$ ($p\approx 6\%$) 
and $N_\uparrow = 505$ ($p\approx 1.4\%$), respectively.
The optimization of the spin coupling 
with respect to both hole doping and interaction energy $U$
is qualitatively similar to that in the RKKY picture. 

\begin{figure}
\vspace*{-70mm}
\hspace*{-38mm}
\psfig{file=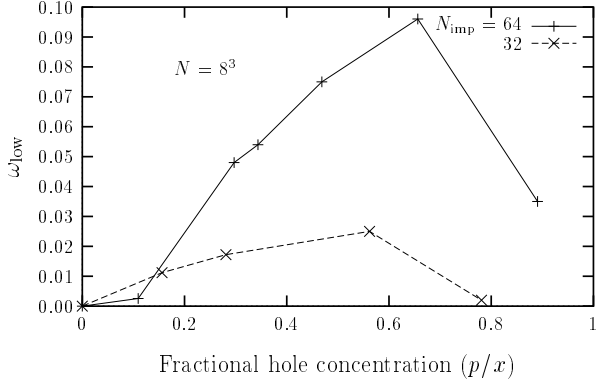,width=135mm,angle=0}
\vspace{-70mm}
\caption{The lowest spin-wave energies 
for the ordered ($+$) and disordered ($\times$) arrangements 
peak near $60\%$ fractional hole concentration,
very similar to the RKKY spin coupling $J^2 \chi_{ij}$ (Fig. 4).
Increasing impurity separation lowers the spin stiffness.}
\end{figure}

\begin{figure}
\vspace*{-70mm}
\hspace*{-38mm}
\psfig{file=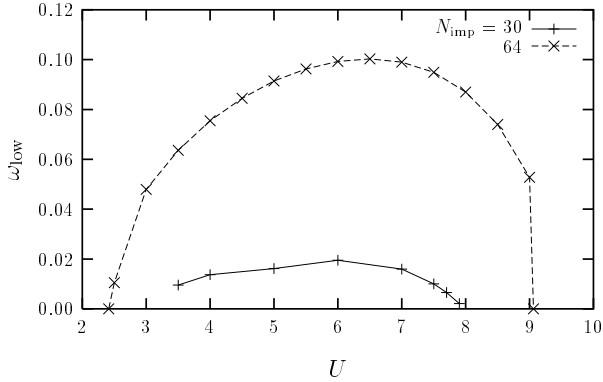,width=135mm,angle=0}
\vspace{-70mm}
\caption{Optimization of the lowest spin-wave energy with $U$, 
qualitatively similar to the generalized RKKY spin coupling (Fig. 7).}
\end{figure}

While the lowest spin-wave energy is softened by disorder, 
the highest spin-wave energy is, however, significantly enhanced,
as shown in Figure 12.
This enhancement of $\omega_{\rm high}$ is associated with localization
of spin-wave states over impurity clusters in which the relatively
closer spins are more strongly coupled.\cite{squid}
For the {\em same} minimum impurity separation ($2a$) in arrangements (i) and (ii),
disorder-induced localization leads to stronger bonds between the cluster spins.
Also shown (for the ordered case) is the Stoner (single-particle excitation) gap,
which is roughly proportional to the MF impurity magnetization. 
The spin-wave branch merges with Stoner excitations at 
about $8\%$ hole concentration.

For the ordered impurity arrangement, 
the spin-wave energy range allows the spin couplings to be extracted, 
as discussed below.
Assuming nearest-neighbour exchange interaction $J$ between the
impurity spins on the superlattice, 
the spin-wave energies are given by
\begin{equation}
\omega_{\bf q}=JSz(1-\gamma_{\bf q})\; ,
\end{equation}

\begin{figure}
\vspace*{-70mm}
\hspace*{-38mm}
\psfig{file=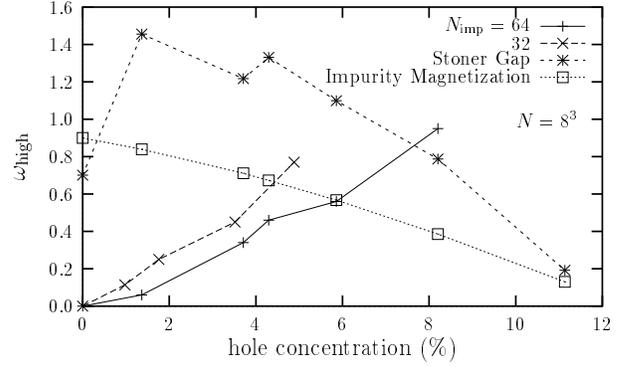,width=135mm,angle=0}
\vspace{-70mm}
\caption{The highest spin-wave energy
for the ordered ($+$) and disordered ($\times$) impurity arrangements,
showing disorder-induced stiffening of the high-energy mode.}
\end{figure}

\noindent
where $\gamma_{\bf q}=(\cos q_x +\cos q_y +\cos q_z)/3$.
The spin-wave modes on the impurity superlattice are plane waves, 
with wave-vector components given by $q_\mu= n_\mu 2\pi/L$, 
where $n_\mu$ are integers and $L=4$ for the 64-impurity superlattice. 
The wave vectors ${\bf q}=(1,0,0)2\pi/L$ etc. and ${\bf q}=(2,2,2)2\pi/L$
correspond to the lowest- and highest-energy modes, respectively. 
The corresponding energies 
$\omega_{\rm low} = JSz/3$ 
and $\omega_{\rm high}=2JSz$ yield a ratio 
$\omega_{\rm high}/\omega_{\rm low} = 6$.
We not only find the actual ratio to be quite close 
(about 7 for most doping cases), 
but the degeneracies in the spin-wave spectrum 
are also in close agreement, 
indicating that nearest-neighbour spin coupling is dominant.

\section{Conclusions}
A comparitive study 
of a generalized RKKY approach and a MF+SF approach offers new and useful
insight into the mechanism of carrier-mediated 
ferromagnetic ordering in a dilute magnetic semiconductor.
While the MF+SF approach provides quantitative understanding of the spin couplings,
competing interactions, spin-wave excitations, low-temperature spin dynamics,
and the critical temperature,
the generalized RKKY approach provides a qualitative understanding in terms of a 
simple physical picture involving the impurity-induced oscillating carrier-spin
polarization, which complements the MF+SF approach.
Our key finding is an optimization of the spin coupling (spin-wave energy) 
with respect to both hole doping and the impurity polarizing field strength ($J$ or $U$),
which is in agreement with recent Monte Carlo study,\cite{mc3}
and can be physically understood in terms of a competition between the increasing 
magnitude of the carrier-spin polarization and increasing rapidity of its oscillation.
We find that the optimum (fractional) hole concentration 
for the spin coupling occurs at $p/x \approx 0.6$, 
and both the spin coupling energy $J^2 \chi(r,J)$ 
and the spin-wave energy scale with the carrier bandwidth $W$,
for fixed $J/W$ or $U/W$.
The oscillating spin polarization also highlights the role of competing interactions
in the instability of the collinear ferromagnetic state.

In this paper, we have presented the first study of spin-wave excitations
in the ferromagnetic state of the DMS within a microscopic correlated lattice fermion
model which treats finite impurity concentration, impurity disorder, and electron 
correlation on an equal footing. 
With regard to electron correlation, the MF+SF approach has been shown 
to be applicable in the full range from weak to strong coupling,\cite{ref1}
and extensively used in the context of strongly correlated layered cuprate 
antiferromagnets which exhibit pronounced spin fluctuations.
When the spin-wave energy is much smaller than the mean-field strength $\Delta$,
spin dynamics is dominant at low temperatures and charge fluctuations can be ignored
for $T<< \Delta$. 
Incorporating the low-energy spin fluctuations about the MF state yields 
quantitatively correct temperature dependence of (sublattice) magnetization
and reliable $T_{\rm C}$ within the renormalized spin-wave-theory (SWT).\cite{ref1}
Whereas $T_{\rm C}$ pertains to global ordering, with the spin coupling energy
providing the relevant energy scale for spin fluctuations, the MFT deals with
local ordering, and greatly over-estimates the transition temperature 
($T_{\rm C} \sim \Delta$), which really represents the moment-melting temperature.

Specifically with regard to the DMS, 
there is a subtle issue concerning the energy scale relevant for global ordering.
Whereas for a generic ferromagnet, 
energy scales corresponding to the local mean field and spin coupling are identical, 
for the DMS, three distinct energies can be identified ---
the two local mean fields seen by the carrier spin ($\sim J$) 
and impurity spin ($\sim J^2 \chi_{ii}$),
and the coupling between impurity spins ($\sim J^2 \chi_{ij}$).
In the weak doping limit ($p/x \rightarrow 0$),
the magnetic response function $\chi(r)$ decays slowly on the impurity-separation
scale, so that $\chi_{ij} \approx \chi_{ii}$,
and the distinction between the two latter energy scales gets blurred.
However, for a realistic fractional doping of $p/x \approx 15\%$,
the impurity spin coupling $J^2 \chi_{ij}$ is by far the lowest energy scale,
and should therefore control the low-temperature behaviour of the magnetization $M(T)$. 
The ferromagnetic transition temperature $T_c$, 
determined within the spin-wave-theory from the
spin coupling energy for a realistic (heavy) hole bandwidth, corresponds closely with
the observed $T_c$ values in $\rm Ga_{1-x}Mn_x As$ and $\rm Ga_{1-x}Mn_x N$.
With appropriate hole doping,
transition temperature much above room temperature appears possible for $x=1/8$,
which is within experimental limit.\cite{limit}

The MF+SF approach also highlights the role of impurity disorder.
While the low-energy spin-wave modes are significantly softened as compared to the 
ordered case, the high-energy spin-wave modes are clearly stiffened,
indicating that a single spin-wave energy scale is not sufficient to describe the 
low-temperature spin dynamics in the DMS. 
In fact, a distribution in spin couplings, with weak and strong bonds, 
has been suggested to be responsible for anomalous temperature dependence
of magnetization, susceptibility, specific heat etc.\cite{anomalous}
Using a simple model involving two spin-excitation energy scales corresponding
to weakly and strongly coupled spins, 
the temperature dependence of magnetization is found to be in good agreement with 
the SQUID magnetization data for Ga$_{1-x}$Mn$_x$As.\cite{squid}

\section*{Acknowledgement}
Helpful discussions with T. Pareek and R. C. Budhani are gratefully acknowledged.

\end{document}